**Terahertz spectra databases require crystallographic information**

Takenori Tanno,[a*] Toru Kurabayashi[b] and Koichi Seo[c]

[a]Venture Incubation Center, Akita University, Akita, Japan

[b]Faculty of Engineering Science, Akita University, Akita, Japan

[c]PANalytical Division, Spectris Co., Ltd., Tokyo, Japan.

*Corresponding author. E-mail: tanno@gipc.akita-u.ac.jp; Tel. & fax: +81-18-889-2803

**Abstract**

Although the sensitivity of THz spectra to the crystal form of the sample being analysed makes it an ideal tool for differentiating polymorphic forms of crystalline materials, the lack of adequate knowledge about the THz response to various materials often results in misinterpretations. The inclusion of structural information in THz spectral databases of crystalline substances is therefore suggested.

**Keywords**

terahertz spectroscopy; far-infrared spectroscopy; database; polymorphism; glycine

Techniques for generating and detecting THz electromagnetic waves have developed rapidly over the last couple of decades, motivated by potential applications in scientific research and information communication. Simultaneously, THz spectroscopy has attracted considerable attention and has been investigated extensively. Characteristic THz absorption spectra of organic substances have been measured and widely reported. Additionally, several reference databases containing THz spectra of various substances exist in the literature and on web sites, e.g., the THz-BRIDGE Spectral Database [1] and the Terahertz Database [2]. However, these reference spectra do take the inherent polymorphism of certain materials into consideration.

Polymorphism is a common characteristic of organic compounds, including both natural substances and pharmaceutical chemicals. The control and assignment of polymorphism is particularly important in pharmacology and in the development of bioactive materials. The polymorphic form of a substance can be controlled or influenced by factors such as the solvent or temperature during nucleation and crystal growth. X-ray diffraction is a reliable method that is used routinely to measure polymorphism and examine the crystal structure of polymorphs. In some cases, a slight shift in the frequency of one or more absorption bands in an infrared spectrum can reveal information about the polymorphism of a substance. Differential scanning calorimetry can also be used to differentiate polymorphs.

Most large molecules can adopt multiple conformations. Differences in molecular skeletal vibrations, intermolecular vibrations, and phonon frequencies strongly reflect differences in molecular conformation. The frequencies of these vibrational modes are in the terahertz (THz) regime. Consequently, it is possible to estimate the polymorphic form of a given substance using THz spectroscopy, as indicated repeatedly in literatures [3] and in the data presented herein.

THz spectra of a polymorphic compound, glycine (Gly) is given in Figure 1. Experimental methods are detailed in Supplementary Information. As expected, the spectra differ considerably. The spectrum of α-Gly contains a large peak at 1.88 THz, whereas the spectrum of γ-Gly contains significant absorption peaks at 2.7, 3.2, 4.2, and 5.2 THz.

$\alpha$-Form glycine molecules are oriented in anti-parallel double layers. In the γ-form, glycine molecules are oriented along the same direction. The hydrogen-bonding network in the molecular crystal and the conformation of each molecule strongly affects the vibrational frequencies observed in THz spectra, and hence, the crystal form of a polymorphic substance can be determined more easily from its THz spectra than from mid- to near-infrared spectra.

For example, the THz spectrum of glycine crystallized from an aqueous solution onto a polyvinylidene fluoride (PVDF) membrane exhibits a steep absorption peak at 1.8-1.9 THz [4]. Therefore, the polymorph that precipitates on the PVDF membrane appears to be α-Gly. This example indicates the merit of THz spectroscopy for polymorph differentiation and the importance of specifying the crystal structure in a reference library.

Despite the usefulness of THz spectroscopy, the polymorphism of any given compound significantly complicates its detection and quantitation. This is particularly important in complex mixtures, and polymorphism must be taken into account when examining the spectra of samples containing a known polymorphic substance.

As mentioned above, several THz spectra databases have been constructed on web, and spectra of compounds, including those of polymorphic substances, have been published. However, crystallographic data for such substances are seldom provided in these databases even in Terahertz Database [2] provided by Riken and NICT (Japan), which is recently updated [5]. Therefore, by referring to these databases without adequate knowledge of polymorphism and its strong relationship with THz-band vibrations, the THz spectrum of a sample may be easily misinterpreted. We recommend that database providers include crystallographic data for all crystalline samples. Additionally, the users of such databases must understand the influence of polymorphism on THz vibrational spectra.

THz spectroscopy is not simply an extension of ordinary IR spectroscopy. THz spectroscopy can be used to access information about the phenomena involved in intermolecular interactions and is not limited to qualitative and quantitative analyses of composition. To develop new frontiers in THz sciences, technologies, and industries, THz spectroscopy needs to be recognised as supramolecular spectroscopy.

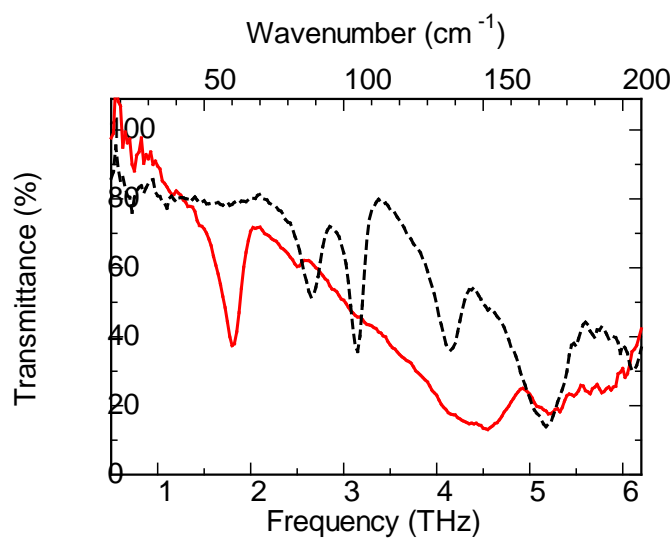

**Fig. 1**  Transmission spectra of α-Gly (solid red line) and γ-Gly (broken black line)